# From Positive- to Negative-Index Materials: Transitional Phenomena

Natalia M. Litchinitser
*Department of Electrical Engineering, The State University of New York at Buffalo,
Buffalo, New York 14260, USA, natashal@buffalo.edu*

Andrei I. Maimistov
*Department of Solid State Physics, Moscow Engineering Physics Institute, Moscow 115409, Russian Federation*

Ildar R. Gabitov
*Department of Mathematics, University of Arizona, Tucson, Arizona 85721, USA*

Roald Z. Sagdeev
*University of Maryland College Park, Maryland, College Park, Maryland 20740, USA*

Vladimir M. Shalaev
*School of Electrical and Computer Engineering and Birck Nanotechnology Center, Purdue University,
West Lafayette, Indiana 47907, USA*

## Abstract

Electromagnetic field propagation through a transition layer between the positive-index and negative-index materials with linearly changing dielectric permittivity and magnetic permeability was investigated. It is shown that at oblique incidence, the components of both TE- and TM-waves exhibit singular behavior in the vicinity of the point where both dielectric permittivity and magnetic permeability are equal to zero. In this case, finite dissipation of electromagnetic field energy takes place even at infinitesimally small values of losses. These results are applicable to a broader class of inhomogeneous metamaterials and may provide a new platform for the realization of low intensity nonlinear optics in metamaterials.

Recent progress in the experimental realization of negative index materials (NIMs) has prompted a reconsideration of many well-known phenomena in electromagnetism. Although negative index of refraction is not found in nature, it was achieved in artificial metamaterials in the microwave [1], terahertz [2], and optical spectral ranges [3-5]. The unusual properties of NIMs most prominently reveal themselves at the interface of positive- and negative-index materials. Combinations of negative- and positive-index materials (PIMs) enable a number of unique physical phenomena and functionalities, including negative refraction [6]; superlensing [7,8]; anti-parallel directions of the phase velocity and the Poynting vector [6,9,10], backward phase matching and negative phase shift facilitating new regimes of second-harmonic generation, parametric processes and bistability [11-18]; unusual surface waves, including surface vortices [19-23] and ultra-compact resonators [24,25]. In most previous studies of spatially combined PIMs and NIMs, sharp transitions from one type of material to the other were considered. The dielectric permittivity $\varepsilon$ and magnetic permeability $\mu$ along with the refractive index $n$ were assumed to be stepwise functions.

One of the most remarkable phenomena that take place at a PIM-NIM interface is that the right-handed triplet of vectors of the electric and magnetic fields and the wave vector $(E, H, k)$ in the PIM undergoes an abrupt change to form a left-handed triplet in the NIM. From a topological viewpoint, a right-handed triplet can not be transformed to a left-handed triplet by any continuous transformations; however, $\varepsilon$ and $\mu$ can be changed from positive to negative values as continuous functions of the coordinate in the direction normal to the interface. The fact that continuously changing material characteristics can lead to a topologically critical phenomenon must be reflected in the transitional characteristics of the electromagnetic waves traversing the interface. In order to understand the physics underlying such a transition, in this letter, we investigate light propagation in a medium consisting of a homogeneous PIM in the region $x<0$ and a homogeneous NIM in the region $x>2h$ separated by the transition layer of width $2h$, where both $\varepsilon$ and $\mu$ are real linear functions of the coordinate $x$ along the normal to the interface such that

$$\varepsilon(x) = \begin{cases} \varepsilon_0 & x<0 \\ \varepsilon_0(1-x/h) & 0<x<2h \\ -\varepsilon_0 & x>2h \end{cases}, \quad \mu(x) = \begin{cases} \mu_0 & x<0 \\ \mu_0(1-x/h) & 0<x<2h \\ -\mu_0 & x>2h \end{cases} \quad (1)$$

as shown in Fig. 1. For simplicity, we neglect herein realistic material losses and Fresnel reflections at interfaces of homogeneous and inhomogeneous regions. It is also assumed that the transition layer width is greater than the wavelength of light $(h>\lambda)$ so that the macroscopic description is valid.

The wave equation for the TE wave ($E$ is perpendicular to the plane of propagation) reads as

$$\frac{\partial^2 E_y}{\partial x^2} + \frac{\partial^2 E_y}{\partial z^2} - \frac{1}{\mu}\frac{\partial \mu}{\partial x}\frac{\partial E_y}{\partial x} + \frac{\omega^2}{c^2}\varepsilon\mu\, E_y = 0, \quad (2)$$

where $E_y$ is an amplitude of the electric component of the harmonic electromagnetic wave at frequency $\omega$ and $c$ is the speed of light in vacuum. The corresponding equation for the TM wave ($H$ is perpendicular to the plane of propagation) can be obtained from Eq. (2) by replacing $E$ with $H$, and $\mu$ with $\varepsilon$, respectively. Here, we only consider the TE case. The components of the magnetic field are related to $E_y$ as follows:

$$H_x = \frac{ic}{\omega\mu}\frac{\partial E_y}{\partial z}, \quad H_z = -\frac{ic}{\omega\mu}\frac{\partial E_y}{\partial x}. \tag{3}$$

Assuming that the medium is homogeneous in z direction, the electric field component can be written as $E_y = \Phi(x)\exp(i\beta z)$. Then, Eq. (2) at $0 < \varsigma < 2$ ($\varsigma = x/h$) takes the following form:

$$\frac{\partial^2 \Phi}{\partial \varsigma^2} + \frac{1}{(1-\varsigma)}\frac{\partial \Phi}{\partial \varsigma} + \left(a^2(1-\varsigma)^2 - b^2\right)\Phi = 0. \tag{4}$$

Here, $a = k_0 h$, $k_0 = (\varepsilon_0 \mu_0)^{1/2}\omega/c$, $b = \beta h$, $\beta = k_0 \sin(\theta_0)$, and $\theta_0$ is the incidence angle. First, we consider the case of normal incidence, that is, $b = 0$. In this case, the solution of Eq. (4) is given by

$$\Phi(\varsigma) = C_1 \exp\left(-ia(1-\varsigma)^2/2\right) + C_2 \exp\left(ia(1-\varsigma)^2/2\right). \tag{5}$$

Expanding $\Phi(\varsigma)$ in the vicinity of $\varsigma = 0$, we obtain

$$\Phi(\varsigma) \approx C_1 \exp(-ia/2 + ia\varsigma + ...) + C_2 \exp(ia/2 - ia\varsigma + ...). \tag{6}$$

The first and second terms in Eq. (6) describe forward and backward propagating waves, respectively. Coefficients $C_1$ and $C_2$ can be found from the boundary conditions at $\varsigma = 0$ and $\varsigma = 2$, taking into account that $\Phi(\varsigma) = A_{in}\exp(ia\varsigma) + A_{ref}\exp(-ia\varsigma)$ if $\varsigma \leq 0$, and $\Phi(\varsigma) = A_{tr}\exp(-ia\varsigma)$ if $\varsigma \geq 2$. Here, $A_{in}$ and $A_{ref}$ are amplitudes of incident and reflected waves respectively, and $A_{tr}$ is the amplitude of the transmitted wave. Straightforward analysis shows that, in the model under consideration, $A_{ref} = 0$ and $C_2 = 0$. Then, the solutions for $E_y$ are given by

$$E_y = \begin{cases} A_{in}\exp(ia\varsigma - i\omega t), & \varsigma \leq 0 \\ A_{in}\exp(-ia(1-\varsigma)^2/2 - i\omega t + ia/2), & 0 < \varsigma < 2 \\ A_{in}\exp(-ia\varsigma - i\omega t + 2ia), & \varsigma \geq 2 \end{cases} \tag{7}$$

Then, using Eq. (3), the solutions for $H_z$ can be written in the form

$$H_z = \begin{cases} A_{in}(\varepsilon_0/\mu_0)^{1/2}\exp(ia\varsigma - i\omega t), & \varsigma \leq 0 \\ A_{in}(\varepsilon_0/\mu_0)^{1/2}\exp(-ia(1-\varsigma)^2/2 - i\omega t + ia/2), & 0 < \varsigma < 2 \\ A_{in}(\varepsilon_0/\mu_0)^{1/2}\exp(-ia\varsigma - i\omega t + 2ia), & \varsigma \geq 2 \end{cases} \quad (8)$$

Figure 2 shows the real part of the electric filed component $E_y$ as a function of $\varsigma$ as the wave propagates from the uniform PIM medium to the uniform NIM medium through a transition layer. In numerical simulations, the spatially dependent material parameters $\varepsilon$ and $\mu$ were taken in the form $\varepsilon(\varsigma) = \varepsilon_0\left(-\tanh\left(\frac{\varsigma-1}{l}\right) + i\delta\right)$ and $\mu(\varsigma) = \mu_0\left(-\tanh\left(\frac{\varsigma-1}{l}\right) + i\delta\right)$, where small loss parameter $\delta = 0.00001$ was introduced to avoid the numerical complications around the point where the real part of $\mu$ approaches zero ($\varsigma = 1$) and $\varepsilon_0 = \mu_0 = 4$. It is noteworthy that the spatial oscillations of the electromagnetic wave are chirped inside the transition layer, suggesting the possibility of a nontrivial behavior of the phase front of the electromagnetic wave within such layer. It follows from Eq. (7) that the phase front is defined by

$$\begin{cases} a\varsigma - \omega t = const & \varsigma < 0 \\ a(1-\varsigma)^2/2 + \omega t = const & 0 < \varsigma < 2; \\ a\varsigma + \omega t = const & \varsigma > 2 \end{cases} \quad (9)$$

therefore, the electromagnetic wave is not a plane wave inside transition layer $0 < \varsigma < 2$. Defining the phase velocity as $v_p = h(d\varsigma/dt) = dx/dt$, we obtain:

$$v_p = \begin{cases} \omega/k_0 & \varsigma < 0 \\ \omega/k_0(1-\varsigma)^{-1} & 0 < \varsigma < 2 \\ -\omega/k_0 & \varsigma > 2 \end{cases} \quad (10)$$

This expression shows that phase velocity has opposite signs at different sides of the point, where both $\varepsilon$ and $\mu$ vanish and $v_p \to \pm\infty$ when $\varsigma \to 1$ from the right (+ sign) and left (− sign) sides, respectively (see Fig. 3).

Next, we consider the case of oblique incidence, that is $b \neq 0$. In this case, Eq. (4) indicates that when $a^2(1-\varsigma)^2 - b^2 < 0$, incoming waves become evanescent. Note that there are two so-called "reflection points" $\varsigma_\pm = 1 \pm b/a$ in our case that are symmetric with respect to $\varsigma = 1$, which makes considered case different from transition of electromagnetic wave through a plasma layer, where there is only one such reflection point, as discussed in Refs. [26-30].

A general solution of Eq. (4) can be written in the form

$$\Phi(\varsigma) = \overline{C}_1 \exp(-ia(1-\varsigma)^2/2) ia(1-\varsigma)^2 U(1-ib^2/4a, 2; ia(1-\varsigma)^2) + $$
$$+ \overline{C}_2 \exp(ia(1-\varsigma)^2/2) ia(1-\varsigma)^2 U(1+ib^2/4a, 2; -ia(1-\varsigma)^2) \equiv \overline{C}_1 \Phi_1 + \overline{C}_2 \Phi_2 \quad (11)$$

where $U$ is a confluent hypergeometric function [31]. Note that the solution (11) reduces to (5) in the case of normal incidence; that is, when $b = 0$. Coefficients $\overline{C}_1$ and $\overline{C}_2$ in Eq. (11) can be determined from boundary conditions assuming that $\Phi(\varsigma) = A_{in} \exp(iqh\varsigma) + A_{ref} \exp(-iqh\varsigma)$ at $\varsigma \leq 0$, and $\Phi(\varsigma) = A_{tr} \exp(-iqh\varsigma)$ at $\varsigma \geq 2$. Here the wave number $q$ is defined as $q^2 = (\omega/c)^2 \varepsilon_0 \mu_0 - \beta^2$.

In order to understand the behavior of the electromagnetic field near the point where both $\mu$ and $\varepsilon$ are changing sign, we expand solutions $\Phi_1$ and $\Phi_2$ in the vicinity of $\varsigma = 1$:

$$\Phi_1(\varsigma) \approx \left(1 + \frac{b^2(1-\varsigma)^2}{4} \ln a(1-\varsigma)^2 - \frac{i}{2}\left(1 - \frac{\pi b^2}{4a}\right) a(1-\varsigma)^2\right),$$
$$\Phi_2(\varsigma) \approx \left(-1 - \frac{b^2(1-\varsigma)^2}{4} \ln a(1-\varsigma)^2 - \frac{i}{2}\left(1 - \frac{\pi b^2}{4a}\right) a(1-\varsigma)^2\right). \quad (13)$$

Then, $\Phi$ takes the form

$$\Phi(\varsigma) \approx (\overline{C}_1 - \overline{C}_2)\left(1 + \frac{b^2(1-\varsigma)^2}{2}\ln\sqrt{a}(1-\varsigma)\right) - \frac{i}{2}(\overline{C}_1 + \overline{C}_2)\left(1 - \frac{\pi b^2}{4a}\right)a(1-\varsigma)^2. \tag{14}$$

The first term in this expression has a logarithmic singularity. The second term corresponds to a regular solution of Eq. (4). The logarithmic term is well defined for $\varsigma < 1$; however, for $\varsigma > 1$, the definition of logarithm is not unique and depends on the choice of the path around $\varsigma = 1$. Following the approach used in Ref. [28], and taking into account the analyticity of $\mu$ in the upper half-plane, we introduce an infinitesimal positive imaginary part (loss) for magnetic permeability and obtain at $\varsigma > 1$:

$$\Phi(\varsigma) \approx (\overline{C}_1 - \overline{C}_2)\left(1 + \frac{b^2(1-\varsigma)^2}{2}\ln\sqrt{a}|1-\varsigma| - i\pi\frac{b^2(1-\varsigma)^2}{2}\right) - \frac{i}{2}(\overline{C}_1 + \overline{C}_2)\left(1 - \frac{\pi b^2}{4a}\right)a(1-\varsigma)^2.$$

$$\tag{15}$$

The additional term $i\pi b^2(1-\varsigma)^2(\overline{C}_1 - \overline{C}_2)/2$ is continuous at $\varsigma = 1$. Taking into account Eq. (3), the definition of $E_y = \Phi(\varsigma)\exp(i\beta z)$, and Eq. (15), we conclude that $E_y$ is finite and continuous at $\varsigma = 1$, while $H_x$ and $H_z$ can be written as

$$\begin{aligned} E_y &= E_0 \exp(i\beta z - i\omega t) + O((1-\varsigma)^2), \\ H_z &= 2i\pi E_0 \varepsilon_0 (h/\lambda)\sin^2\theta_0 \ln\sqrt{a}(1-\varsigma)\exp(i\beta z - i\omega t) + O((1-\varsigma)^2), \\ H_x &= -E_0\sqrt{\varepsilon_0/\mu_0}(1-\varsigma)^{-1}\sin\theta_0 \exp(i\beta z - i\omega t) + O((1-\varsigma)^2), \end{aligned} \tag{16}$$

where $\theta_0$ is the incidence angle and $E_0 = \overline{C}_1 - \overline{C}_2$. Thus, the $x$ component of the magnetic field $H_x$ is singular at $\varsigma = 1$, and the $z$ component $H_z$ experiences a jump when the value of $\mu$ changes sign from positive to negative at the point $\varsigma = 1$. Following [28], we find that the difference of the longitudinal components of the Poynting vector $S_x = (c/4\pi)E_y H_z$ (averaged over rapid field oscillations) before and after the transition of the wave through the point $\varsigma = 1$ is given by $\Delta S_x = c\pi E_0^2 (h/\lambda)\varepsilon_0 \sin^2\theta_0$. Note that, while in our original model we assumed losses (imaginary parts of $\varepsilon$ and $\mu$) to be infinitesimally small, at the point $\varsigma = 1$, the real parts of $\varepsilon$ and $\mu$ are zero, and therefore, the contribution of the small imaginary parts of $\varepsilon$ and $\mu$ becomes significant and can no longer be neglected. Using the technique proposed in Ref. [28], we calculate the dissipation of energy $\Delta Q$ due to these losses at $\varsigma = 1$, and find that $\Delta Q = \Delta S_x$. Therefore, transition through the point where $\varepsilon$ and $\mu$ are changing sign even with infinitesimal dissipation is accompanied by a finite loss of incident wave energy.

Figure 4(a) illustrates the real part of the electric field component $E_y$ as a function of a longitudinal coordinate $\varsigma$ for a fixed angle of incidence and fixed width of the transition layer. Figure 4(b) shows $|H_x|$ versus $\varsigma$ confirming the prediction that $H_x$ becomes infinite at $\varsigma = 1$. While we have not assumed any specific model for the inhomogeneous metamaterial, in a simplified way the origin of the anomalous field enhancement shown in Fig. 4 can be understood in a simplified way as a spatial analog of a well-known resonance occurring in a spectral domain when, for example, light interacts with a harmonic oscillator [27]. It is noteworthy that since in the vicinity of the $\varepsilon = 0$ (and $\mu = 0$) point the wavelength of light becomes very large, the system can effectively be considered as quasistatic [32]. In the case of the TM wave, the thin layer near the $\varepsilon = 0$ point can be considered as a very thin capacitor that accumulates infinitely large electric field energy, if we neglect the effects of dissipation and spatial dispersion. Note that such energy accumulation occurs only for obliquely incident waves since the electric field at the oblique incidence has a non-zero component in the direction of propagation. Since an electric displacement D must be continuous, the electric field E anomalously increases as $\varepsilon$ tends to zero. Likewise, for the TE wave (considered here), the magnetic field has a non-zero component in the direction of propagation, and the magnetic field energy accumulates in the vicinity of the $\mu = 0$ point in space. Such a thin layer near the $\mu = 0$ point can be considered as a short solenoid that stores the magnetic field energy. In this case, H anomalously increases as $\mu$ tends to zero. Finally, owing to the singularities of the magnetic field components, the $x$- and $z$-components of the Poynting vector are also singular at $\varsigma = 1$. In addition, that $z$-component of the Poynting vector, given by $S_z = -(c/4\pi)E_y H_x$, is changing sign passing the point $\varsigma = 1$. This corresponds to a change of sign of the refraction angle at $\varsigma = 1$.

Considered model represents a significant simplification of real metamaterials. Clearly, electromagnetic field components (and the components of the Poynting vector) tending to infinity do not represent a physical solution, but result from the assumption of lossless media. Losses can be included by adding non-negligible imaginary parts in both $\varepsilon$ and $\mu$. The effect of loss is particularly important in the context of NIMs, since all practically realized NIMs to date are significantly lossy. The field enhancement factors in the presence of realistic losses, the angular dependence, and the area where the field enhancement occurs are of significant fundamental and practical interest and will be discussed in a longer publication. Finally, it should be mentioned that, in realistic metamaterials, the real parts of $\varepsilon$ and $\mu$ are crossing zero at different spatial points, and the imaginary parts $\varepsilon$ and $\mu$ inside the transition layer are spatially dependent. Although we only considered the case of the TE-wave in this paper, similar effects can be predicted for the TM-wave case.

To summarize, we analyzed electromagnetic wave propagation through the PIM-NIM transition layer of finite width with linearly changing $\varepsilon$ and $\mu$. Several unique features of light transmission through this inhomogeneous layer were found: (i) at normal incidence, phase velocity tends to infinity in the middle of the layer where $\varepsilon = 0$ and $\mu = 0$, and it changes sign past this point; (ii) the electric and magnetic field components are finite and contain no singularities at normal incidence; however, spatial oscillations are chirped in the vicinity of the $\varepsilon = 0$ and $\mu = 0$ point; (iii) at oblique incidence in the case of the TE-wave, the electric field component is continuous and finite, while one of the magnetic field components tends to infinity as $1/(1-\varsigma)$ in a lossless case, and another component contains a logarithmic singularity; (iv) at oblique incidence, energy dissipation takes place during wave transition through the point $\varsigma = 1$, even if material losses are infinitesimally small.

It is noteworthy that similar effects have been discussed previously in a context of inhomogeneous plasma, with the dielectric permittivity changing as a function of the longitudinal coordinate [26-28]. However, in the plasma case, the field enhancement near the $\varepsilon = 0$ point was only predicted for the TM-wave case (since $\mu = 1$). Also, since in that case there are no propagating waves supported past the point $\varepsilon = 0$ (where $\varepsilon < 0$), the fields evanescently decay beyond that point.

The phenomena predicted here may be applied to a wider class of inhomogeneous metamaterials. Field enhancement effects in the vicinity of the $\varepsilon = 0$ and $\mu = 0$ point present entirely new opportunities for the realization of nonlinear optics at low input intensities and antenna applications [29,30]. Owing to the great flexibility of the design of the material parameters (or of the index of refraction) enabled by metamaterial technology, the experimental realization of the effects predicted in this study should be feasible in the microwave region and in optics when bulk NIMs become available.

The authors would like to thank V. G. Veselago, J. B. Pendry, R. W. Boyd, A. M. Rubenchik, V. E. Zakharov, A. C. Newell, G. Shvets, A. O. Korotkevich, E. V. Kazantseva, and

T. I. Lakoba for enlightening discussions. This research was supported by the Army Research Office through grants W911NF-07-1-034 and 50342-PH-MUR, by the National Science Foundation through grant DMS-050989, and by the Russian Foundation for Basic Research through grant 06-02-16406.**References**

[1] R. A. Shelby, D. R. Smith, and S. Schultz, Science **292**, 77 (2001).

[2] S. Linden, C. Enkrich, M. Wegener, J. Zhou, T. Koschny, and C. M. Soukoulis, Magnetic response of metamaterials at 100 Terahertz, Science **306**, 1351 (2004).

[3] V. M. Shalaev, W. Cai, U. K. Chettiar, H. Yuan, A. K. Sarychev, V. P. Drachev, and A.V. Kildishev, Opt. Lett. **30**, 3356 (2005).

[4] S. Zhang, W. Fan, N. C. Panoiu, K. J. Malloy, R. M. Osgood, and S.R. J. Brueck, Phys. Rev. Lett. **95**, 137404 (2005).

[5] G. Dolling, C. Enkrich, M. Wegener, C. M. Soukoulis, and S. Linden, Opt. Lett. **31**, 1800 (2006).

[6] V. G. Veselago, Soviet Physics Uspekhi **10**, 509 (1968).

[7] J. B. Pendry, Phys. Rev. Lett. **85**, 3966 (2000).

[8] Z. Liu, N. Fang, T.-J. Yen, and X. Zhang, Appl. Phys. Lett. **83**, 5184 (2003).

[9] L. I. Mandelshtam, *Complete Collected Works* (Akad. Nauk SSSR, Moscow, 1947).

[10] V. M. Agranovich and Yu. N. Gartstein, Usp. Fiz. Nauk **176**, 1051 (2006).

[11] V. M. Agranovich, Y. R. Shen, R. H. Baughman, and A. A. Zakhidov, Phys. Rev. B **69**, 165112 (2004).

[12] A. K. Popov and V. M. Shalaev, Opt. Lett. **31**, 2169 (2006).

[13] A. K. Popov and V. M. Shalaev, Appl. Phys. B **84**, 131 (2006).
11

**Captions**

Figure 1. A schematic of a transition layer between the PIM and the NIM with linearly changing dielectric permittivity and magnetic permeability.

Figure 2. The real part of the electric field component $E_y$ as a function of longitudinal coordinate $\varsigma$ for the case of normal incidence (solid line), the real part of the magnetic susceptibility $\mu$ as a function of longitudinal coordinate $\varsigma$ (long-dashed line), and the boundary between the PIM and the NIM (short-dashed line).

Figure 3. Normalized phase velocity $v_p k_0/\omega$ as a function of normalized longitudinal coordinate $\varsigma$ outside and within the transition layer. Long-dashed line shows the boundary between the PIM and the NIM, and short-dashed lines show the limits of the transition layer.

Figure 4. (a) The real part of the electric field component $E_y$ as a function of longitudinal coordinate $\varsigma$ for the case of oblique incidence at $\theta = \pi/17$ (solid line), the real part of the magnetic susceptibility $\mu$ as a function of longitudinal coordinate $\varsigma$ (long-dashed line), and the boundary between the PIM and the NIM (short-dashed line), (b) The absolute value of the normalized magnetic field component $|H_x|$ as a function of normalized longitudinal coordinate $\varsigma$ (solid line) and the boundary between the PIM and the NIM (short-dashed line).

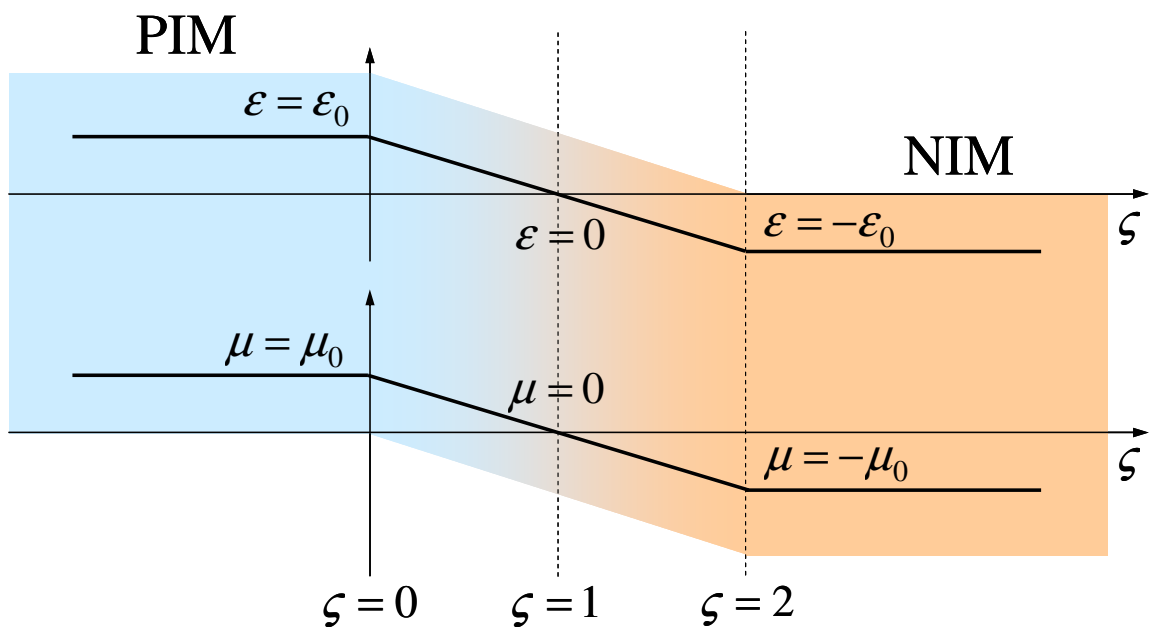

Figure 1

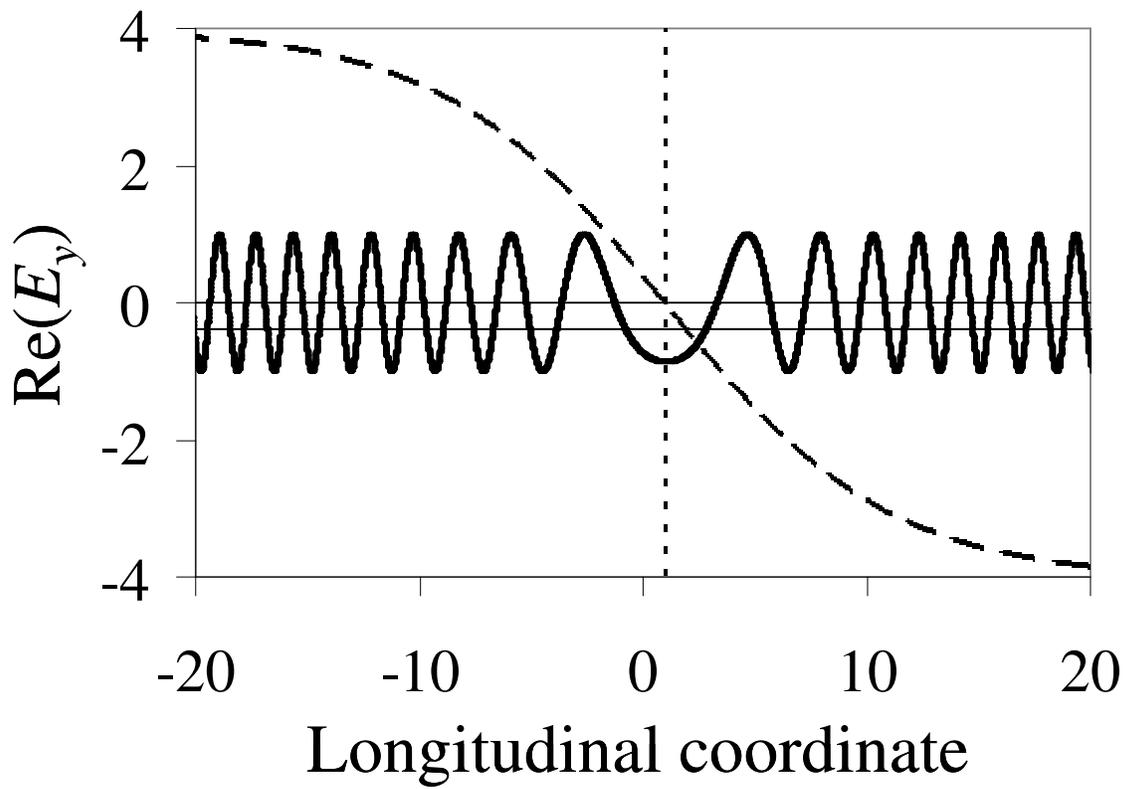

Figure 2

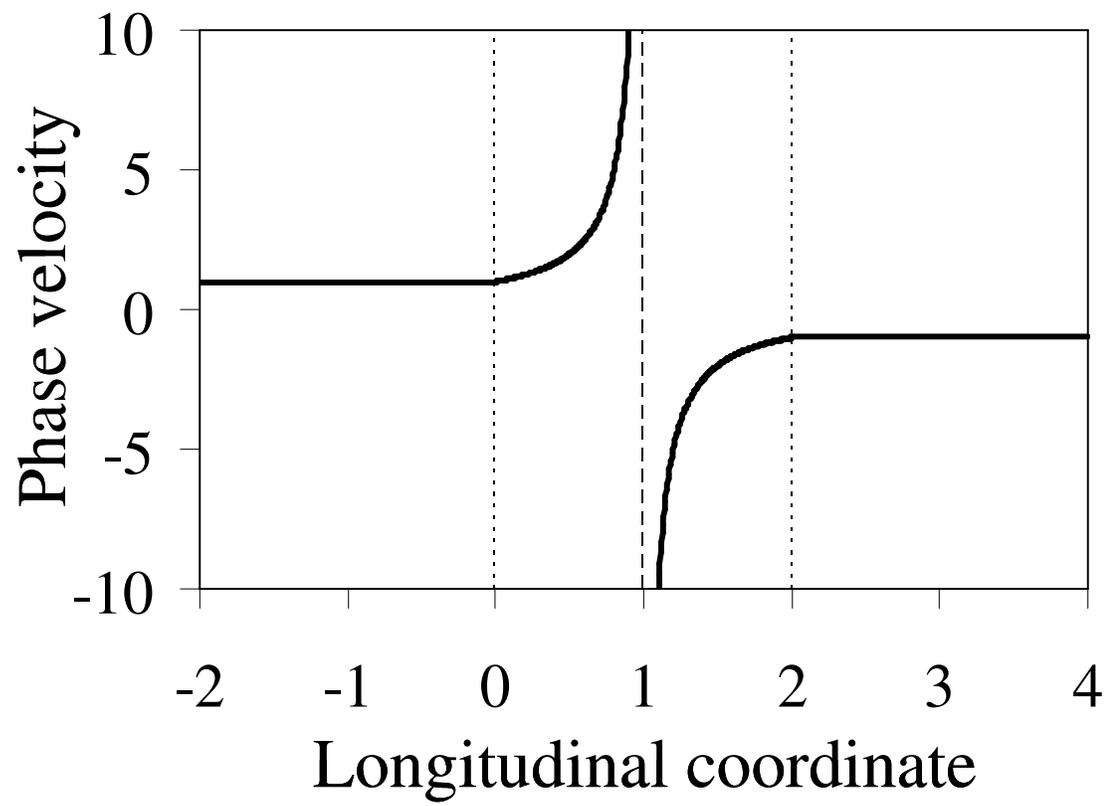

Figure 3

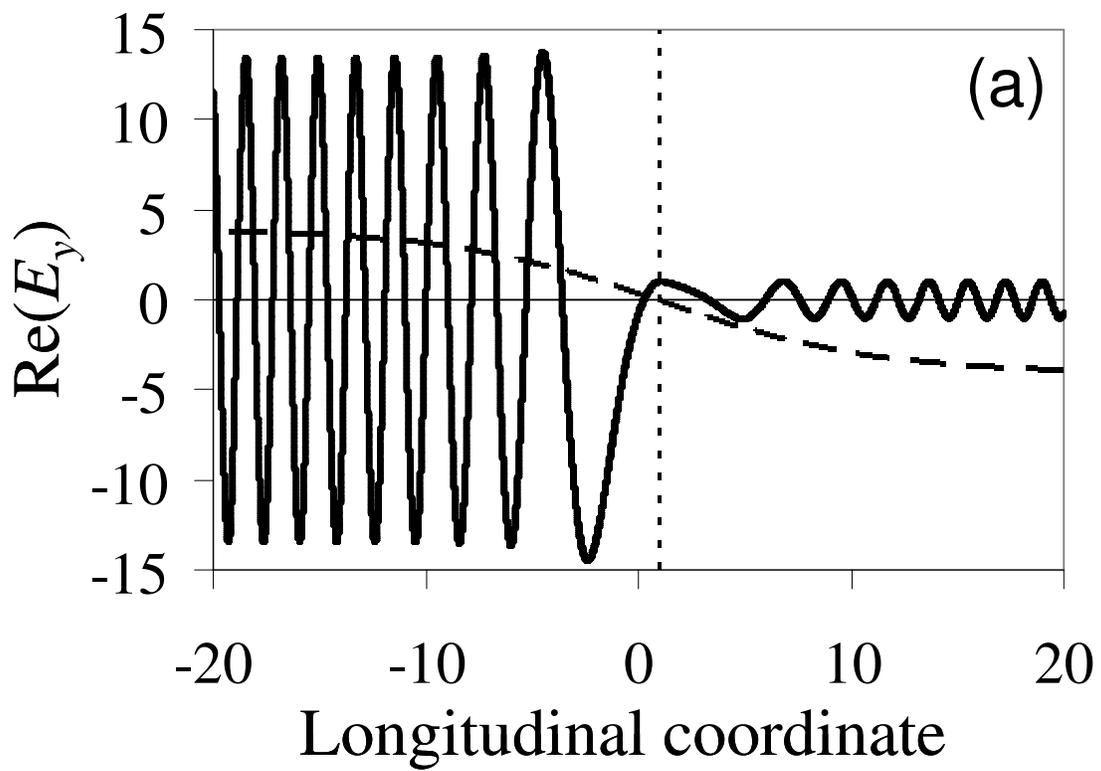

Figure 4(a)

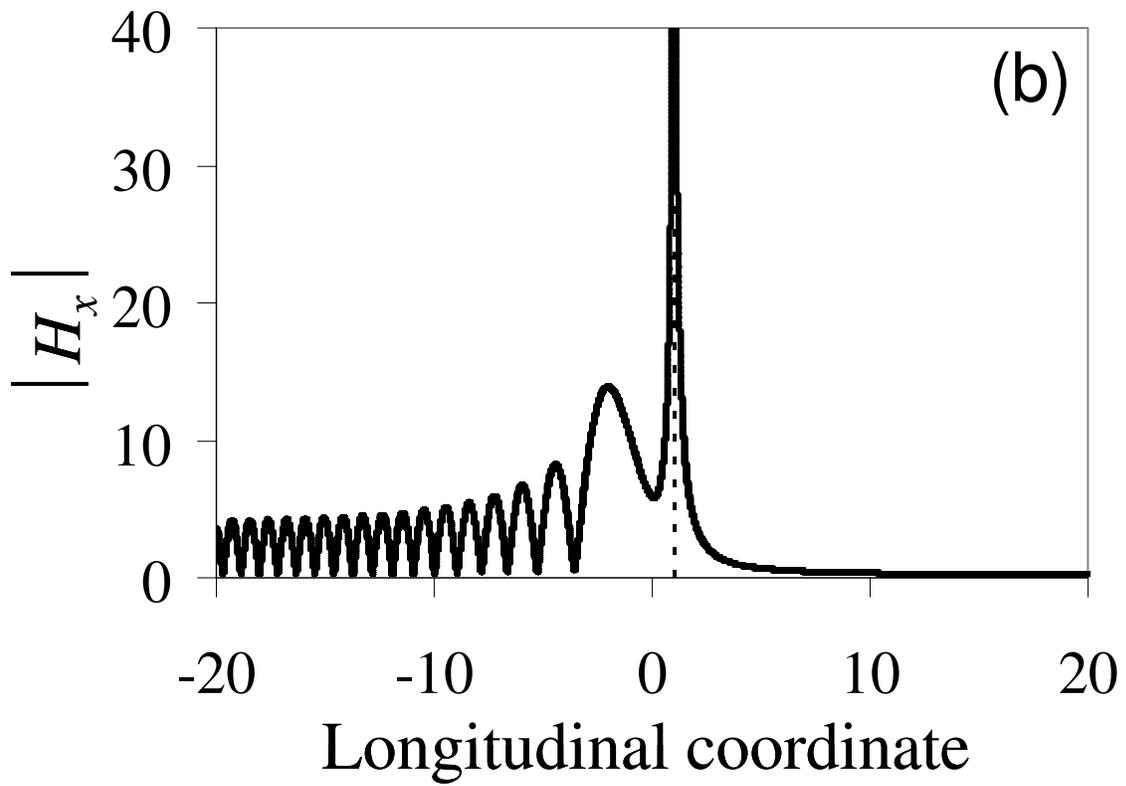

Figure 4(b)